\documentclass[
showpacs,
aps,
prb,
twocolumn,
superscriptaddress,
10pt,
longbibliography
]{revtex4-1}

\usepackage{graphicx}
\usepackage{bm}
\usepackage{color}
\usepackage[colorlinks=true, allcolors=blue]{hyperref}
\usepackage{enumerate}
\usepackage{xspace}
\usepackage{mathrsfs}
\usepackage{mathptmx}
\usepackage[utf8]{inputenc}
\usepackage{enumitem}
\usepackage{amssymb}
\usepackage{amsmath}
\newcommand{\angstrom}{\text{\normalfont\AA}}
\usepackage{listings}
\usepackage{float}
\usepackage{color}
\usepackage{multirow}
\usepackage{subfigure}
\usepackage[table]{xcolor}
\usepackage{tabularx}
\usepackage{soul}
\usepackage{siunitx}
  \usepackage{natbib}

\begin{document}

\title{Modeling coupled spin and lattice dynamics}

\author{Mara Strungaru}
\affiliation{Department of Physics, University of York, York, United Kingdom}

\author{Matthew O A Ellis}
\affiliation{Department of Computer Science, University of Sheffield, Sheffield, United Kingdom} 

\author{Sergiu Ruta}
\affiliation{Department of Physics,University of York, York, United Kingdom}

\author{Oksana Chubykalo-Fesenko}
\affiliation{Instituto de Ciencia de Materiales de Madrid, CSIC, Madrid, Spain}

\author{Richard F L Evans}
\affiliation{Department of Physics, University of York, York, United Kingdom}

\author{Roy W Chantrell}
\affiliation{Department of Physics, University of York, York, United Kingdom}

\begin{abstract}
A unified model of molecular and atomistic spin dynamics is presented enabling simulations both in microcanonical and canonical ensembles without the necessity of additional phenomenological spin damping. Transfer of energy and angular momentum between the lattice and the spin systems is achieved by a coupling term based upon the spin-orbit interaction. The characteristic spectra of the spin and phonon systems are analyzed for different coupling strength and temperatures. The spin spectral density shows magnon modes together 
with the uncorrelated noise induced by the coupling to the lattice. The  effective damping parameter  is investigated showing an increase with both coupling strength and temperature. The model paves the way to understanding magnetic relaxation processes beyond the phenomenological approach of the Gilbert damping and  the dynamics of the  energy transfer between lattice and spins. 
\end{abstract}

\maketitle
\section{Introduction}

With the emergent field of ultrafast magnetisation dynamics \cite{beaurepaire1996ultrafast} understanding the flow of energy and angular momentum between electrons, spins and phonons  is crucial for the interpretation of the wide range of observed phenomena~\cite{Dornes2019,Walowski2016,Radu2015,Hennecke2019}. For example, phonons strongly pumped in the THz regime by laser excitation can modulate the exchange field and manipulate the magnetisation as shown for the magnetic insulator YIG ~\cite{maehrlein2018dissecting} or in Gd~\cite{Melnikov2003}. The excitation of THz phonons leads to a magnetic response with the same frequency in Gd~\cite{Melnikov2003}, proving the necessity of considering the dynamics of both lattice and spins. Phonon excitations can modulate both anisotropy and exchange which can successfully manipulate ~\cite{kim2015controlling, Kim2012, Scherbakov2010} or potentially switch the magnetisation ~\cite{Vlasov2020,Kovalenko2013}, ultimately leading to the development of low-dissipative memories.

Magnetisation relaxation is typically modeled using the phenomenological description of damping proposed by Landau and Lifshitz \cite{LLeq} and later Gilbert   \cite{gilbert2004phenomenological}, where the precessional equation of motion is augmented by a friction-like term, resulting in the Landau-Lifshitz-Gilbert (LLG) equation. This represents the coupling of the magnetic modes (given primarily by the localised atomic spin) with the non-magnetic modes (lattice vibrations and electron orbits). The LLG equation and its generalisations can be
deduced from the quantum-mechanical approaches assuming an equilibrium phonon bath and the weak coupling of the spin to the bath degrees of freedom \cite{Garanin1991,Rebei2003, Nieves2014}.  Thus the standard approach works on the supposition that
the time scales between the environmental degrees of freedom and the magnetic degrees of freedom are well separated and reducing the coupling between the magnetization and its environment  to a single phenomenological damping parameter \cite{ellis2015landau, bhattacharjee2012atomistic}. In reality, the lattice and magnetisation dynamics have comparable time-scales, where the interaction between the two subsystems represents a  source of damping, hence the necessity of treating spin and lattice dynamics in a self-consistent way.

To investigate these phenomena, and aiming at predictive power for the design of competitive ultrafast magnetic nano-devices, advanced frameworks beyond conventional micromagnetics and atomistic spin dynamics\cite{evans2014atomistic} are needed \cite{Dieny2020}. A complete description of magnetic systems includes the interaction between several degrees of freedom, such as lattice, spins and electrons, modeled in a self-consistent simulation framework. The characteristic relaxation timescales of electrons are much smaller ($\approx \si{fs}$) in comparison to spin and lattice ($\SI{100}{fs}-\si{ps}$), hence magnetisation relaxation processes can be described via coupled spin and lattice dynamics, termed Spin-Lattice Dynamics (SLD) ~\cite{Beaujouan2012,Ma2008, Ma2009,Tranchida2018,Perera2016,Assmann2019,Hellsvik2019,Karakurt2007}. SLD models can be crucial in disentangling the interplay between these sub-systems.

SLD models have so far considered either micro-canonical (NVE - constant particle number, volume and energy) \cite{Assmann2019,Hellsvik2019} or canonical (NVT - constant particle number, volume and temperature) ensembles with two Langevin thermostats connected to both lattice and spin subsystems~\cite{Ma2012, Ma2008}. Damping due to spin-lattice interactions only within the canonical ensemble (NVT) has not yet been addressed, but is of interest in future modelling of magnetic insulators at finite temperature. Here we introduce a SLD model capable of describing both ensembles. Specifically, our model (i) takes into account the transfer of angular momentum from spin to lattice and vice-versa, (ii) works both in a micro-canonical ensemble (constant energy) and in a canonical ensemble (constant temperature), (iii) allows a fixed Curie temperature of the system independent of the spin-lattice coupling strength, (iv) disables uniform translational motion of the system and additional constant energy drift, which can be produced by certain spin-lattice coupling forms. Furthermore, in this work, the characteristics of the induced spin-lattice noise, the magnon-phonon induced damping and the equilibrium properties of the magnetic system has been systematically investigated. 

The paper is organised as follows. We start by describing the computational model of Spin-Lattice Dynamics and the magnetic and mechanical energy terms used in this framework (Section \ref{sec::comp_mod}). We then explore the equilibrium properties of the system for both microcanonical and canonical simulations, proving that our model is able to efficiently transfer both energy and angular momentum between the spin and lattice degrees of freedom. In Section \ref{sec::eq_prop} we compute the equilibrium magnetisation as function of temperature for both a dynamic and static lattice and we show that the order parameter is not dependent on the details of the thermostat used. In Section \ref{sec::corr} we analyse the auto-correlation functions and spectral characteristics of magnon, phonons and the coupling term proving that the pseudo-dipolar coupling efficiently mediates the transfer of energy from spins to the lattice and vice-versa. We then calculate the temperature and coupling dependence of the induced magnon-phonon damping and we conclude that the values agree well with damping measured in magnetic insulators, where the electronic contributions to the damping can be neglected ( Section \ref{sec::magnon_phonon_damping}).

\section{Computational model}
\label {sec::comp_mod}
In order to model the effects of both lattice and spin dynamics in magnetic materials an atomistic system is adopted with localised atomic magnetic moments at the atomic coordinates. Within this framework there are now nine degrees of freedom; atomic magnetic moment (or spin) $\mathbf{S}$, atomic position $\mathbf{r}$ and velocity $\mathbf{v}$.
The lattice and the magnetic system can directly interact with each other via the position and spin dependent Hamiltonians. The total Hamiltonian of the system consists of a lattice $\mathcal{H}_{lat}$ and magnetic  $\mathcal{H}_{mag}$ parts: 
\begin{equation}
\mathcal{H}_{\mathrm{tot}}= \mathcal{H}_{\mathrm{lat}} + \mathcal{H}_{\mathrm{mag}} .
\end{equation}

The lattice Hamiltonian includes the classical kinetic and pairwise inter-atomic potential energies:
\begin{equation}
\mathcal{H}_{\mathrm{lat}}= \sum_{i} \frac {m_i \textbf{v}_i^2} {2} + \frac{1}{2} \sum_{i,j}U(r_{ij}) .
\end{equation}

Our model considers a harmonic potential (HP) defined as:
\begin{equation}
 U(r_{ij})= \begin{cases}
  {V_0} (r_{ij}-r_{ij}^0)^2/ a_0^{2} & r_{ij}<r_c \\
 0 & r_{ij}>r_c .
 \end{cases}
\end{equation}
where $V_0$ has been parametrised for \textsc{bcc} Fe in \cite{Assmann2019} and  $a_0=1\angstrom$ is a dimension scale factor. To be more specific we consider the equilibrium distances $r_{ij}^0$ corresponding to a symmetric \textsc{bcc} structure. The interaction cut-off is $r_c=\SI{7.8}{\angstrom}$. The parameters of the potential are given in Table \ref{parameters_sld}. The harmonic potential has been used for simplicity, however it can lead to rather stiff lattice for a large interaction cutoff.

Another choice of the potential used in our model is an anharmonic Morse potential (MP) parameterised in \cite{girifalco1959application} for \textsc{bcc} Fe and defined as:

\begin{equation}
U(r_{ij})= \begin{cases}
D [e^{ -2 a (r_{ij}-r_0)}- 2 e^{-a (r_{ij}-r_0)}] & r_{ij}<r_c \\
0 & r_{ij}>r_c
\end{cases}
\end{equation}

The Morse potential approximates well the experimental phonon dispersion observed experimentally for \textsc{bcc} Fe \cite{minkiewicz1967phonon} as shown in \cite{ellis2015simulations}. The phonon spectra for the choices of potential used in this work are given in Section \ref{sec::corr}.
Other nonlinear choices of potential can be calculated via the embedded atom method \cite{dudarev2005magnetic,derlet2007million}. 

The spin Hamiltonian ($\mathcal{H}_{\mathrm{mag}}$) used in our simulations consists of contributions from the exchange interaction, Zeeman energy and a spin-lattice coupling Hamiltonian, given by the  pseudo-dipolar coupling term ($\mathcal{H}_{\mathrm{c}}$), which we will describe later:

\begin{equation}
\mathcal{H}_{\mathrm{mag}}= - \frac{1}{2}\sum_{i,j} J(r_{ij})(\textbf{S}_i \cdot \textbf{S}_j) - \sum_{i}\mu_i \mathbf{S}_i \cdot \mathbf{H}_\text{app} + \mathcal{H}_{\mathrm{c}} ,
\end{equation}

where $\mu_i$ is the magnetic moment of atom $i$,  $\textbf{S}_{i}$ is a unit vector describing its spin direction  and $\mathbf{H}_\text{app}$ is an external applied magnetic field.
The  exchange interactions used in our simulations depend on atomic separation $J(r_{ij})$. They were  calculated from first principle methods for \textsc{bcc} Fe  by Ma et al \cite{Ma2008} and follow the dependence:
\begin{equation}
J(r_{ij}) = J_0 \left(1- \frac {r_{ij}} {r_c} \right)^3 \Theta (r_c-r_{ij}),
\end{equation}
where $r_c$ is the cutoff and $\Theta(r_c-r_{ij})$ is the Heaviside step function, which implies no exchange coupling between spins situated at larger distance than $r_c$.

Several previous SLD models suffered from the fact that they did not allow angular momentum transfer between lattice and spin systems   \cite{Hellsvik2019}. This happened for magnetisation dynamics in the absence of spin thermostat, governed by symmetric exchange only, due to total angular momentum conservation. To enable transfer of angular momentum, Perera \textit{et al}~\cite{Perera2016} have proposed local anisotropy terms to mimic the spin-orbit coupling phenomenon due to symmetry breaking of the local environment. Their approach was successful in thermalising the subsystems, however, single site anisotropy spin terms with a position dependent coefficients as employed in~\cite{Perera2016} can induce an artificial collective translational motion of the sample while the system is magnetically saturated, due to the  force $-\frac {\partial H_{\mathrm{tot}}} {\partial \mathbf{r}_i}$ proportional to spin orientation. To avoid large collective motion of the atoms in the magnetic saturated state, we consider a two-site coupling term, commonly known as the pseudo-dipolar coupling, described by
 
\begin{equation}
\label{eq::coupling}
\mathcal{H}_c= -\sum_{i,j} f({r}_{ij}) \left[(\textbf{S}_i \cdot \hat {\textbf{r}}_{ij}) (\textbf{S}_j \cdot \hat {\textbf{r}}_{ij}) - \frac {1}{3} \textbf{S}_i \cdot \textbf{S}_j \right].
\end{equation}
 The origin of this term still lies in the spin-orbit interaction, appearing from the dynamic crystal field that affects the electronic orbitals and spin states. It  has been employed previously in SLD simulations ~\cite{Beaujouan2012, Assmann2019}. It was initially proposed by Akhiezer~\cite{akhiezer1968spin}, having the same structure of a dipolar interaction, however with a distance dependence that falls off  rapidly, hence the name pseudo-dipolar interaction. The exchange-like term $-\frac {1}{3} \textbf{S}_i \cdot \textbf{S}_j$ is necessary in order to preserve the Curie temperature of the system under different coupling strengths and to ensure no net anisotropy when the atoms form a symmetric cubic lattice. For the mechanical forces, the exchange like term  eliminates the anisotropic force that leads to a non-physical uniform translation of the system when the magnetic system is saturated.  The magnitude of the interactions is assumed to decay as $f(\textbf{r}_{ij})=C J_0 (a_0 /r_{ij})^4$ as presented in~\cite{Assmann2019} with $C$ taken as a constant, for simplicity measured relative to the exchange interactions and $a_0=1\angstrom$ is a dimension scale factor. The constant $C$ can be estimated from ab-initio calculations \cite{Perera2016}, approximated from magneto-elastic coefficients \cite{Assmann2019}, or can be chosen to match the relaxation times and damping values, as in this work. 

Since the total Hamiltonian now depends on the coupled spin and lattice degrees of freedom ($\textbf{v}_i$, $\textbf{r}_i$, $\textbf{S}_i$), the following equations of motion (EOM) need to be solved concurrently to obtain the dynamics of our coupled system:

\begin{align}
\frac{\partial \textbf{r}_i}{\partial t} &= \textbf{v}_i , & & \\
\frac{\partial\textbf{v}_i}{\partial t} &= - \eta \mathbf{v}_i + \frac {\textbf{F}_i}{m_i}, \\
\frac{\partial \mathbf{S}_i }{\partial t} & = -{\gamma} \mathbf{S}_i \times \mathbf{H}_i, \\
\mathbf{F}_i & = -\frac {\partial \mathcal{H}_{tot}} {\partial \mathbf{r}_i} +{\Gamma}_i ,\\ 
\mathbf{H}_i &=- \frac{1}{\mu_S \mu_0}\frac {\partial \mathcal{H}_{tot}} {\partial \textbf{S}_i},
\end{align} 
where $\mathbf{F}_i$ and $\mathbf{H}_i$ represent the effective force and field, $ \boldmath{\Gamma}_i $ represents the fluctuation term (thermal force) and $\eta$ represents the friction term that controls the dissipation of energy from the lattice into the external thermal reservoir. The strength of the fluctuation term can be calculated by converting the dissipation equations into a Fokker-Planck equation and then calculating the stationary solution. The thermal force has the form of a Gaussian noise:
\begin{align}
\langle \Gamma_{i\alpha}(t) \rangle & =0, \\
\langle \Gamma_{i\alpha}(t)\Gamma_{j\beta}(t') \rangle &= \frac {2 \eta k_B T}{m_i} \delta_{\alpha\beta} \delta_{ij} \delta(t-t') .
\label{th_noise}
\end{align}

To prove that the complete interacting many-body spin-lattice framework presented in here is in agreement with the fluctuation-dissipation theorem, we have followed the approach presented by Chubykalo \textit{et al}~\cite{chubykalo2003brownian} based on the Onsager relations. Linearising the equation of motion for spins, we find that the kinetic coefficients for the spin system are zero, due to the fact that the spin and internal field are thermodynamic conjugate variables. Hence, if the noise applied to the lattice obeys the fluctuation dissipation theory, the coupled system will obey it as well, due to the precessional form of the equation of motion for the spin.

We compare the SLD model presented here with other existing model that do not take into account the lattice degrees of freedom (Atomistic Spin Dynamics - ASD). Particularly, in our case we assume a fixed lattice positions. The summary of the comparison is presented in Table \ref{summary_compare_models}. Atomistic spin dynamics simulations (ASD)~\cite{evans2014atomistic,eriksson2017atomistic,ellis2015landau,spirit2009} have been widely used to study finite size effects, ultrafast magnetisation dynamics and numerous other magnetic phenomena. Here the  intrinsic spin damping (the Gilbert damping -  $\alpha_{\mathrm{G}}$) is phenomenologically included.  In our case since the lattice is fixed it is assumed to come from electronic contributions. Consequently, only 3 equations of motion per atom describing the spin dynamics are used:

\begin{align}
 \frac{\partial \mathbf{S}_i }{\partial t}   = -\frac{\gamma}{(1+\alpha_{\mathrm{G}}^2) } \mathbf{S}_i \times  (\mathbf{H}_{i}+   \alpha_{\mathrm{G}} \mathbf{S}_i \times \mathbf{H}_{i} )   
 \label{llg}
\end{align}

with an additional field coming from the coupling to the fixed lattice positions. The temperature effects are introduced in spin variables by means of a Langevin thermostat. The spin thermostat is modeled by augmenting the effective fields by a thermal stochastic field ($\mathbf{H}_i = \boldsymbol{\xi}_i - \partial \mathcal{H}/\partial \mathbf{S}_i$) and its properties also follow the fluctuation-dissipation theorem:
\begin{align}
\langle \xi_{i\alpha}(t) \rangle &=0, \\
\langle \xi_{i\alpha}(t)\xi_{j\beta}(t') \rangle &= \frac {2  \alpha_{\mathrm{G}} k_B T }{\gamma \mu_S} \delta_{\alpha,\beta} \delta_{ij} \delta(t-t') .
\label{th}
\end{align}

The characteristics of the above presented models are summarised in Table \ref{summary_compare_models}.

\begin{table}
\centering
\fontsize{9}{8}
\begin{ruledtabular}
\begin{tabular}{c|l|l|l|l}
Model~ & Lattice~ & Lattice~  &  Spin~  & Intrinsic Spin
 \\ 
    &     & thermostat~  &  thermostat~ &damping~
 \\\hline
SLD & Dynamic & On & Off& Phonon \\
 & & & & induced \\ \hline
ASD & Fixed & Off & On& Electronic \\
 & & & & mainly \\
\end{tabular}
\end{ruledtabular}
\caption{Summary comparison of the SLD model developed here against other spin dynamics models. }
\label{summary_compare_models}
\end{table}

\begin{table}\centering
\fontsize{9}{8}
\begin{ruledtabular}
\begin{tabular}{c|lll}
Quantity & Symbol & Value  &  Units  
 \\ \hline \hline 
Exchange \cite{Ma2008} & $J_0$ & $0.904$  & eV    \\
~ & $r_c$ & $3.75$  & $\angstrom$    \\ \hline
Harmonic potential \cite{Assmann2019} & $V_0$ & $0.15$  & eV    \\
~& $r_c$ & $7.8$  & $\angstrom$    \\ \hline
Morse potential \cite{girifalco1959application} & $D$ & $0.4174$  & eV    \\
~& $a$ & $1.3885$  & $\angstrom$    \\ 
~& $r_0$ & $2.845$  & $\angstrom$    \\ 
~& $r_c$ & $7.8$  & $\angstrom$    \\ \hline
Magnetic moment & $\mu_s$ & $2.22$  & $\mu_B$    \\
Coupling constant & $C$ & $0.5 $  & $~$    \\
Mass & $m$ & $55.845$  & $u$    \\
Lattice constant& $a$ & $2.87$  & $\angstrom$    \\ 
Lattice damping & $\eta$ & $0.6$  & s$^{-1}$    \\
\end{tabular}
\end{ruledtabular}
\caption{Parameters used in the spin-lattice model to simulate \textsc{bcc} Fe.}
\label{parameters_sld}
\end{table}

 To integrate the coupled spin and lattice equations of motion we used a Suzuki-Trotter decomposition (STD) method \cite{suzuki1976generalized} known for its numerical accuracy and stability. The scheme can integrate non-commuting operators, such as is the case of spin-lattice models and conserves the energy and space-phase volume. The conservation of energy is necessary when dealing with microcanonical simulations. Considering the generalized coordinate $ \textbf{X}=\{ \textbf{r} , \textbf{v}, \textbf{S} \}$ the equations of motion can be re-written using the Liouville operators:
\begin{align}
\frac{\partial \textbf{X}}{\partial t}&= \hat{L} \textbf{X(t)}=(\hat{L}_r+\hat{L}_v+\hat{L}_S)\textbf{X}(t). 
\end{align} 
The solution for the Liouville equation is $\textbf{X}(t+\Delta t)= e ^ {L \Delta t} \textbf{
X}(t)$. Hence, following the form of this solution and applying a Suzuki-Trotter decomposition as in Tsai's work \cite{Tsai2005, Tsai2004}, we can write the solution as: 
\begin{align}
\textbf{X}(t+\Delta t) &= e^{\hat{L}_s\frac{\Delta t}{2}} e^{\hat{L}_v\frac{\Delta t}{2}} e^{\hat{L}_r\Delta t} e^{\hat{L}_v\frac{\Delta t}{2}} e^{\hat{L}_s\frac{\Delta t}{2}} \textbf{X}(t) + O(\Delta t^3),
\end{align} 
where $L_s, L_v, L_r$ are the Liouville operators for the spin, velocity and position. This update can be abbreviated as $\textbf{(s,v,r,v,s)}$ update. 
The velocity and position are updated using a first order update, however the spin needs to be updated using a Cayley transform \cite{omelyan2001algorithm,lewis2003geometric}, due to the fact that the norm of each individual spin needs to be conserved. Thus we have 
\begin{align}
e^{\hat{L}_v\Delta t} \mathbf{v}_i&=\mathbf{v}_i+ \frac {\Delta t}{m_i}\mathbf{F}_i ,\\
e^{\hat{L}_r\Delta t} \mathbf{r}_i&=\mathbf{r}_i+ \Delta t \mathbf{v}_i ,\\
e^{\hat{L}_S\Delta t} \mathbf{S}_i&= \frac {\mathbf{S}_i + \Delta t \mathbf{H}_i \times \mathbf{S}_i + \frac {\Delta t^2}{2} \left[(\mathbf{H}_i \cdot \mathbf{S}_i)\mathbf{H}_i- \frac{1}{2}\mathbf{H}_i^2 \mathbf{S}_i \right]} { 1+ \frac{1}{4}{\Delta t^2}\mathbf{H}_i^2 }.
\end{align} 

\par
The spin equations of motions depend directly on the neighbouring spin orientations (through the effective field) hence individual spins do not commute with each other. We need to further decompose the spin system $\hat{L}_s=\sum_i \hat{L}_{s_i}$. The following decomposition will be applied for the spin system:

\begin{align}
e^{\hat{L}_s(\Delta t/2)} = e^{\hat{L}_{s_1}(\Delta t/4)}...e^{\hat{L}_{s_N}(\Delta t/2)}...e^{\hat{L}_{s_1}(\Delta t/4)} + O(\Delta t^3)
\end{align} 

Tests of the accuracy of the integration have
been performed by checking the conservation of energy within the microcanonical ensemble. 
To ensure that the spin and lattice sub-systems have reached equilibrium, we  calculate both the lattice temperature (from the Equipartition Theorem) and spin temperature \cite{ma2010temperature}. These are defined as:

\begin{align}
T_L= \frac{2}{3Nk_B} \sum_i \frac {\mathbf{p}_i^2}{2m}, ~~~ T_S= \frac{\sum_i (\mathbf{S}_i \times \mathbf{H}_i)^2} {2 k_B \sum_i{\mathbf{S}_i \cdot \mathbf{H}_i}}.
\end{align}

\section{Spin-lattice thermalisation}
\label{sec::eq_prop}

As an initial test of our model we look at the thermalisation process  within micro-canonical (NVE) and canonical (NVT) simulations for a periodic \textsc{bcc} Fe system of $10 \times 10 \times 10$ unit cells. No thermostat is applied directly to the spin system and its thermalisation occurs via transfer of energy and angular momentum from the lattice, i.e. via the magnon-phonon interaction. In the case of the NVE simulations, the energy is deposited into the lattice by randomly displacing the atoms from an equilibrium \textsc{bcc} structure positions within a \SI{0.01}{\angstrom} radius sphere and by initialising their velocities with a Boltzmann distribution at $T = 300$~K. The spin system is initialised randomly in the $x-y$ plane with a constant component of magnetisation of 0.5 in the out of plane ($z$) direction. In the case of NVT simulations, the lattice is connected to a thermostat at a temperature of $T=300$~K. The parameters used in the simulations are presented in Table \ref{parameters_sld}.

\begin{figure}[tb]
 \centering   
 \includegraphics[width=\columnwidth, trim={2cm 2cm 0 0}]{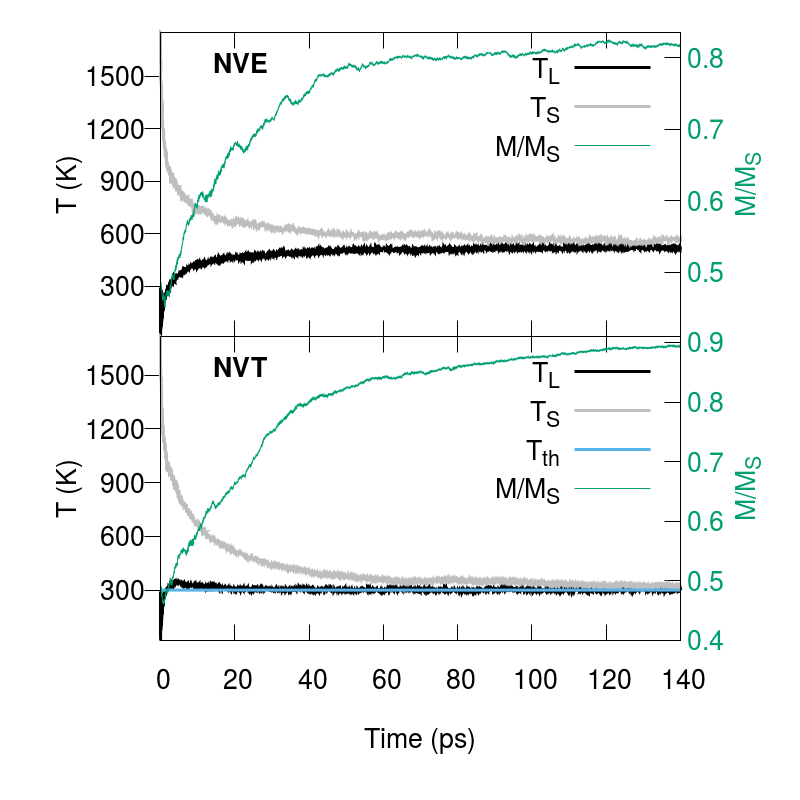}
 \caption{ NVE (top) and NVT (bottom) simulations for a $10 \times 10 \times 10$ unit cell \textsc{bcc} Fe system. The spin system is randomly initialised with a temperature of 1900~K, while the lattice velocities are initialised by a Boltzmann distribution at $T=300$~K. In both cases we obtain equilibration of the two subsystems on the ps timescale. }{\label{nve_nvt}}
\end{figure}

Fig. \ref{nve_nvt} shows the thermalisation process for the two types of simulation. In both cases the spin system has an initial temperature of $T=1900$~K due to the random initialisation. For the NVE simulations, the two subsystems are seen to equilibrate at a temperature of $T=600$~K, this temperature being dependent on the energy  initially deposited into the system. In the NVT simulations, the lattice is thermalised at $T = 300$~K followed by the relaxation of the spin towards the same temperature. In both cases we observe that the relaxation of the spin system happens on a \SI{100}{ps} timescale, corresponding to typical values for spin-orbit relaxation. The corresponding change in the magnetisation is emphasized by the green lines in Fig. \ref{nve_nvt} showing a transfer of angular momentum between the spin and lattice degrees of freedom.

To gain a better understanding of properties at thermal equilibrium within the Spin-Lattice Dynamics model, we have investigated the temperature dependence of the magnetic order parameter in different frameworks that either enable or disable lattice dynamics, specifically: SLD or ASD. Tab.~\ref{summary_compare_models} illustrates the differences between the models. Since reaching joint thermal equilibrium depends strongly on the randomness already present in the magnetic system this process is accelerated by starting with a reduced magnetisation of $M/M_{\mathrm{S}}=0.9$ for $T>300$~K.

 \begin{figure}[tb]\centering     
 \includegraphics[width=\columnwidth, trim={2cm 1cm 0 0}]{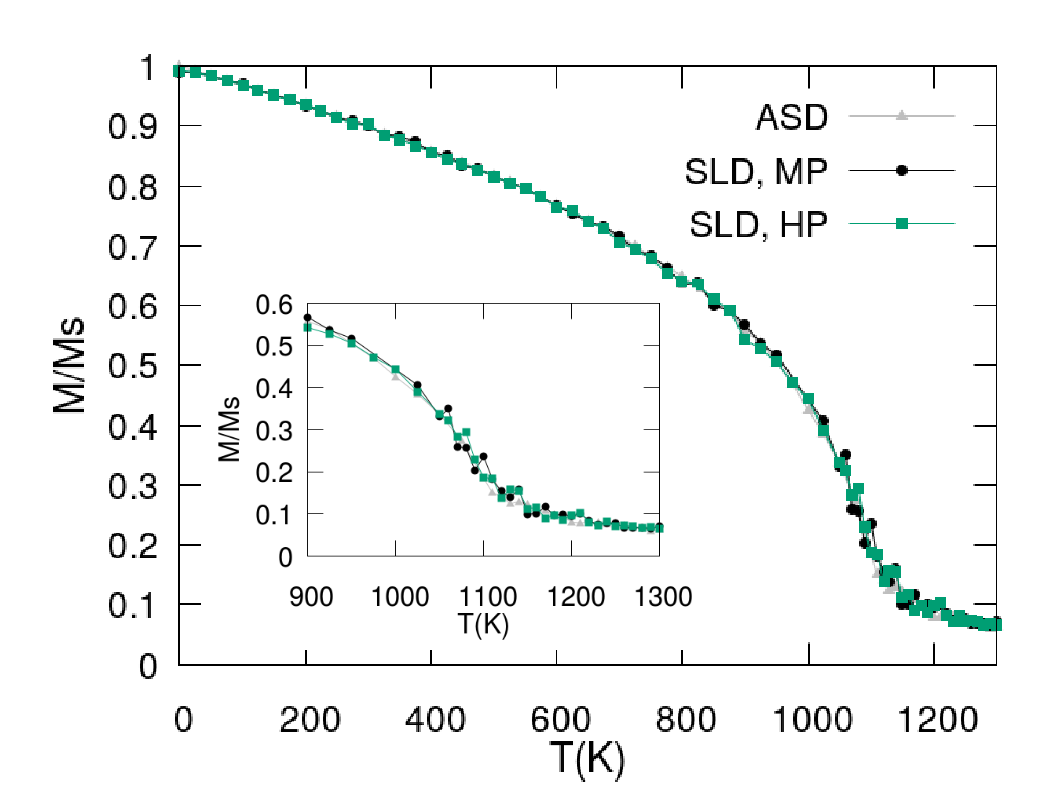}
 \caption{Magnetisation versus temperature curves for the SLD model (with different choices of lattice potential: MP-Morse Potential, HP-Harmonic Potential) and fixed lattice ASD model. The inset zooms around the ferromagnetic to paramagnetic phase transition temperature.}{\label{mvst_harm_morse}}
\end{figure}

Fig. \ref{mvst_harm_morse} shows the comparison of the equilibrium magnetisation using either the harmonic potential (HP), Morse potential (MP) or fixed lattice (ASD) simulations. The magnetisation is calculated by averaging for 200 ps after an initial equilibration for 800 ps (for SLD type simulations) or 100 ps (for ASD) simulations. We observe that even without a spin thermostat (in SLD model) the  magnetisation reaches equilibrium via the thermal fluctuations of the lattice, proving that both energy and angular momentum can be successfully transferred between the two sub-systems. Additionally, both the SLD and ASD methods give the same equilibrium magnetisation over the temperature range considered. This confirms that the equilibrium quantities are independent of the details of the thermostat used, in agreement with the fact that both SLD and ASD models obey the fluctuation-dissipation theorem.

In principle, since the strength of the exchange interaction depends on the relative separation between the atoms, any thermal expansion of the lattice could potentially modify the Curie temperature. However, as highlighted in the inset of Fig.~\ref{mvst_harm_morse}, the same Curie temperature is observed in each model. We attribute this to fact that the thermal lattice expansion is small in the temperature range considered due to two reasons: i) the Curie temperature of the system is well below the melting point of Fe ($\approx \SI{1800}{K}$) and ii) we model a bulk, constant-volume system with periodic boundary conditions that does not present strong lattice displacements due to surfaces. We note that Evans \textit{et al}~\cite{evans2006influence} found a reduction of $T_{\mathrm{C}}$ in nanoparticles due to an expansion of atomic separations at the surface  
 that consequently reduces the exchange interactions. For systems with periodic boundary conditions we anticipate fluctuations in the exchange parameter due to changes in interatomic spacings to be relatively small. Although the equilibrium properties are not dependent on the details of the thermostat or the potential, the magnetisation dynamics could be strongly influenced by these choices. 

\begin{figure}[tb]\centering   
 \includegraphics[width=1\columnwidth, trim=0.2in 1in 0in 0in]{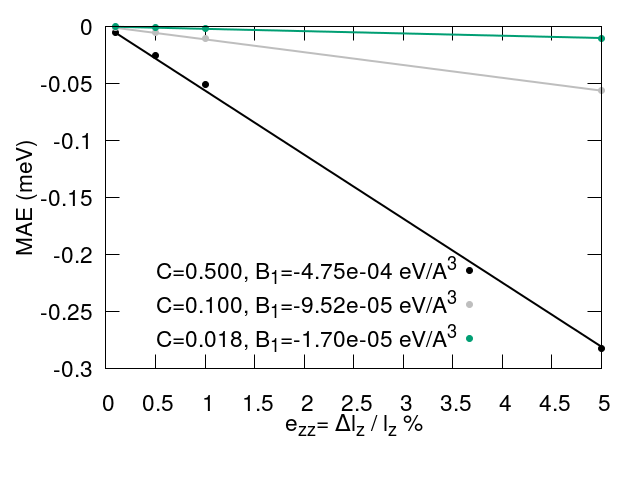}
 \caption{Magnetic anisotropy energy as function of strain for different coupling strengths for T=0K.}{\label{fig:magnetoelastic}}
\end{figure}

The strength of the pseudo-dipolar coupling parameters $C$ determines the timescale of the thermalisation process. Its value can be parametrised from magneto-elastic simulations via calculations of the anisotropy energy as a function of strain. The magneto-elastic Hamiltonian can be written for a continuous magnetisation $\textbf{M}$ and elastic strain tensor $\textbf{e}$ as \cite{kittel1949physical,kamra2015coherent}:
\begin{equation}
\label{eq::me_hamilt}
\mathcal{H}_{\mathrm{m-e}}= \frac{B_1}{M_{\mathrm{S}}^2}\sum_{i} M_i^2 e_{ii} + \frac{B_2}{M_s^2}\sum_{i} M_i M_j e_{ij}
\end{equation}
where constants $B_1,B_2$ can be measured experimentally \cite{Sander1999}. The pseudo-dipolar term acts as a local anisotropy, however, for a lattice distorted randomly, this effective anisotropy is averaged out to zero. At the same time under external strain effects, an effective anisotropy will arise due to the pseudo-dipolar coupling which is the origin of the magneto-elastic effects. To calculate the induced magnetic anisotropy energy (MAE), the \textsc{bcc} lattice is strained along the $z$ direction whilst fixed in the $xy$ plane. The sample is then uniformly rotated and the energy barrier is evaluated from the angular dependence of the energy. Fig.~\ref{fig:magnetoelastic} shows MAE for different strain values and coupling strengths, with the magneto-elastic energy densities constants $B_1$ obtained from the linear fit. The values of the obtained constants $B_1$ are larger than the typical values reported for \textsc{bcc} Fe $B_1=-3.43$ MJ m$^{-3}=-6.2415 \times 10^{-6}$ eV A$^{-3}$ \cite{Sander1999} measured at $T=300$~K. Although the obtained magneto-elastic coupling constants for \textsc{bcc} Fe are larger than experimental values, it is important to stress that, as we will see later, a large coupling is necessary in order to obtain  damping parameters comparable to the ones known for magnetic insulators where the main contribution comes from magnon-phonon scattering. In reality, in \textsc{bcc} Fe there is an important contribution to the effective damping from electronic sources, which if considered, can lead to the  smaller coupling strengths, consistent in magnitude with experimental magneto-elastic parameters. Indeed, as we will show later, our finding suggests that phonon damping is a very small contribution in  metallic systems such as \textsc{bcc} Fe .

\section{Dynamic Properties at Thermal Equilibrium}
\label{sec::corr}

\begin{figure*}[!tb]\centering     
 \includegraphics[width=1.5\columnwidth, trim={2cm 3cm 0 0}]{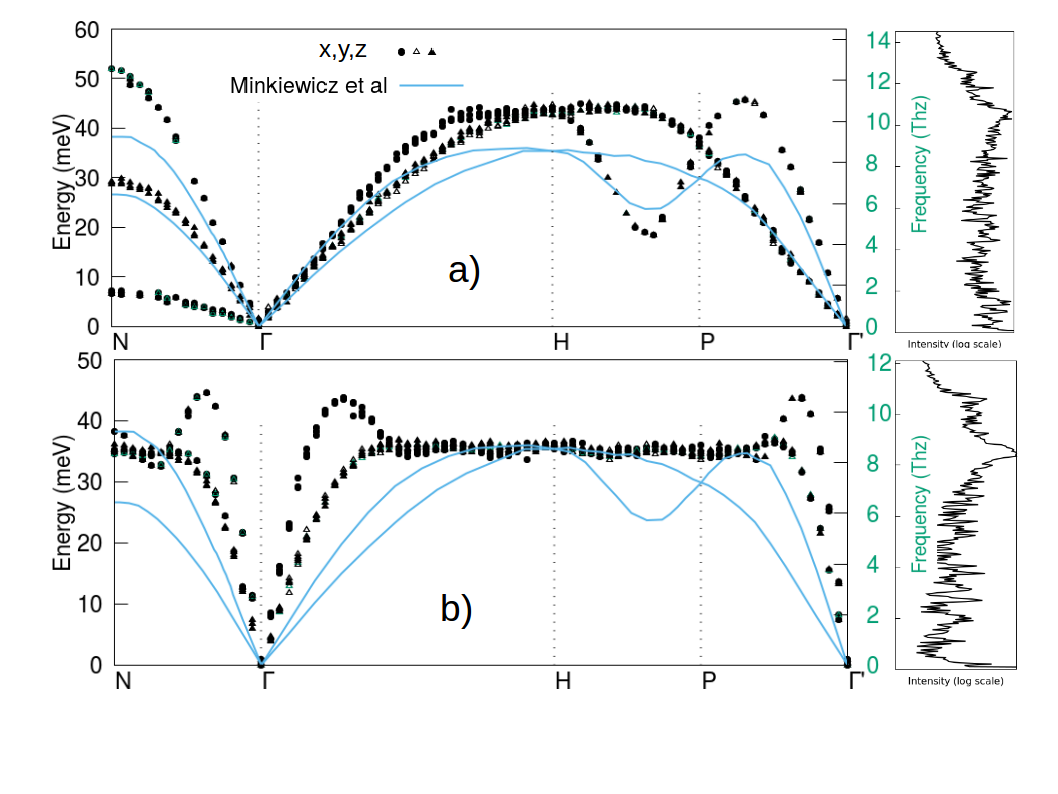}
 \caption{Phonon spectra calculated for a $32 \times 32 \times 32$ unit cell system at $T=300$K, $C=0.5$ for a) Morse potential, b)Harmonic potential. The spectra are calculated via method $i)$.} Right figure includes the projection of the intensity of the spectra onto the frequency domain. Solid lines are the experimental data of Minkiewicz \textit{et al}~\cite{minkiewicz1967phonon}. For the Minkiewicz \textit{et al} data there is only 1 datapoint for the N-$\Gamma$ path for the second transverse mode which does not show up on the line plots. {\label{phonons_spectra}}
\end{figure*}

\begin{figure}[!htb]\centering     
 \includegraphics[width=1\columnwidth, trim={2cm 12cm 0 0}]{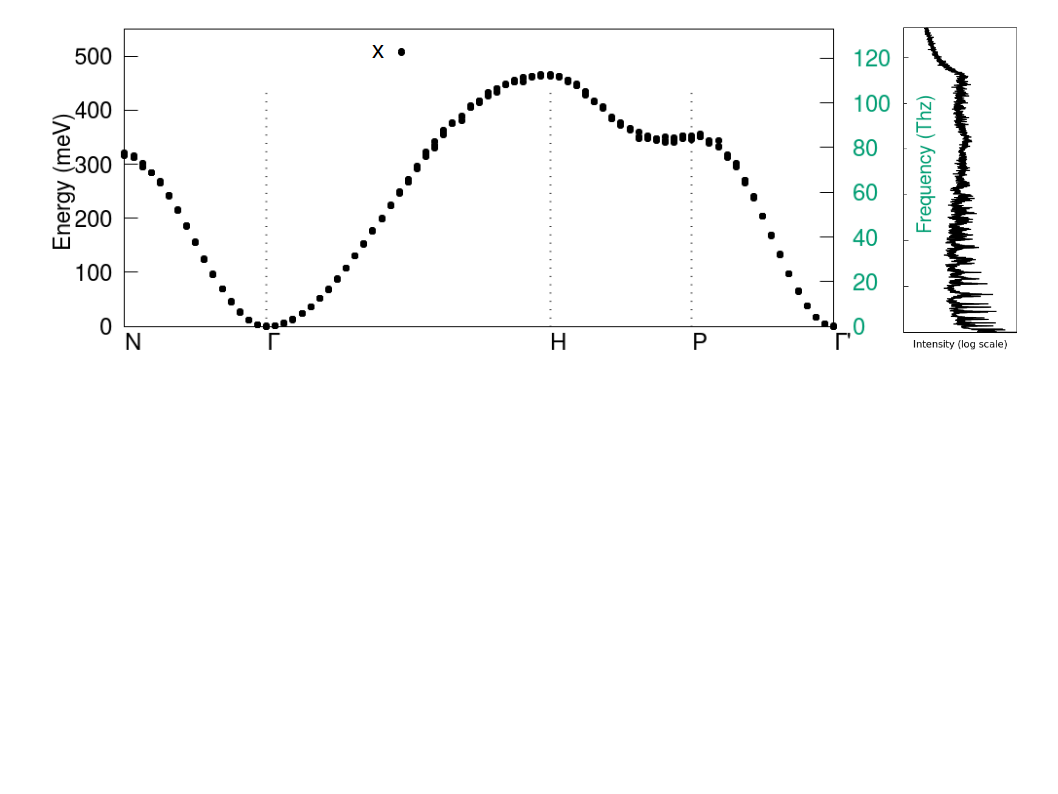}
 \caption{Magnon spectrum (x component) calculated for a $32 \times 32 \times 32$ unit cell system at $T=300$K, $C=0.5 $ for a Morse potential. The spectrum is calculated via method $i)$.} Right figure includes the projection of the intensity of the spectrum onto the frequency domain.  {\label{magnons_morse_spectra}}
\end{figure}

Section \ref{sec::eq_prop} showed that the equilibrium magnetisation does not depend on the details of the thermostat used and a successful transfer of both energy and angular momentum is achieved between the spin and lattice sub-systems by the introduction of a pseudo-dipolar coupling term. In this section, we investigate the properties of the magnons, phonons and the coupling term that equilibrates the spin and phonon systems in the absence of a phenomenological spin damping. Two types of simulations are presented here: $i)$ magnon and phonon spectra calculated along the high symmetry path of a \textsc{bcc} lattice and $ii)$ averaged temporal Fourier transform (FT) of individual atoms datasets (spin, velocity, pseudo-dipolar coupling field). The phonon  - Fig. \ref{phonons_spectra} and magnon - Fig. \ref{magnons_morse_spectra} spectra are calculated by initially equilibrating the system for 10 ps with a spin thermostat with $ \alpha_{\mathrm{G}}=0.01$ and a coupling of $C=0.5$, followed by 10 ps of equilibration in the absence of a spin thermostat.  For the method $i)$ the correlations are computed for a runtime of 20 ps after the above thermalisation stage. For each point in $k$-space, the first three maxima of the auto-correlation function are plotted for better visualisation. The auto-correlation function is projected onto the frequency space and the average intensity is plotted for different frequencies. The phonon spectra are calculated from the velocity auto-correlation function defined in Fourier space as~\cite{papanicolaou1995modification,ellis2015simulations}:

\begin{equation}
  A^p(k,\omega)= \int_0^{t_f} \langle v_k^p(t)v_k^p(t-t') \rangle e^{-i \omega t } dt
\end{equation}
where $p=x,y,z$, $t_f$ is the total time and $v_k^p(t)$ is the spatial Fourier Transform calculated numerically as a discrete Fourier Transform:
\begin{equation}
  v_k^p(t)=\sum_i v_i^p e^{-i \mathbf{k} \cdot \mathbf{r_i}}
\end{equation}
The same approach is applied for the magnon spectra, using the dynamical spin structure factor, which is given by the space-time Fourier transform of the spin-spin correlation function defined as $C^{mn} (r-r', t-t')=<S^m(r,t)S^n(r',t')>$, with $m,n$ given by the \textit{x,y,z} components \cite{krech1998fast}:

\begin{equation}
  S^{m n}(\mathbf{k},\omega)= \sum_{r,r'} e^{i \mathbf{k} \cdot (\mathbf{r}-\mathbf{r'})} \int_0^{t_f} C^{m n} (r-r', t-t') e^{-i \omega t } dt 
\end{equation}

\begin{figure*}[t]\centering     
 \includegraphics[width=2\columnwidth, trim={2cm 10cm 2cm 0}]{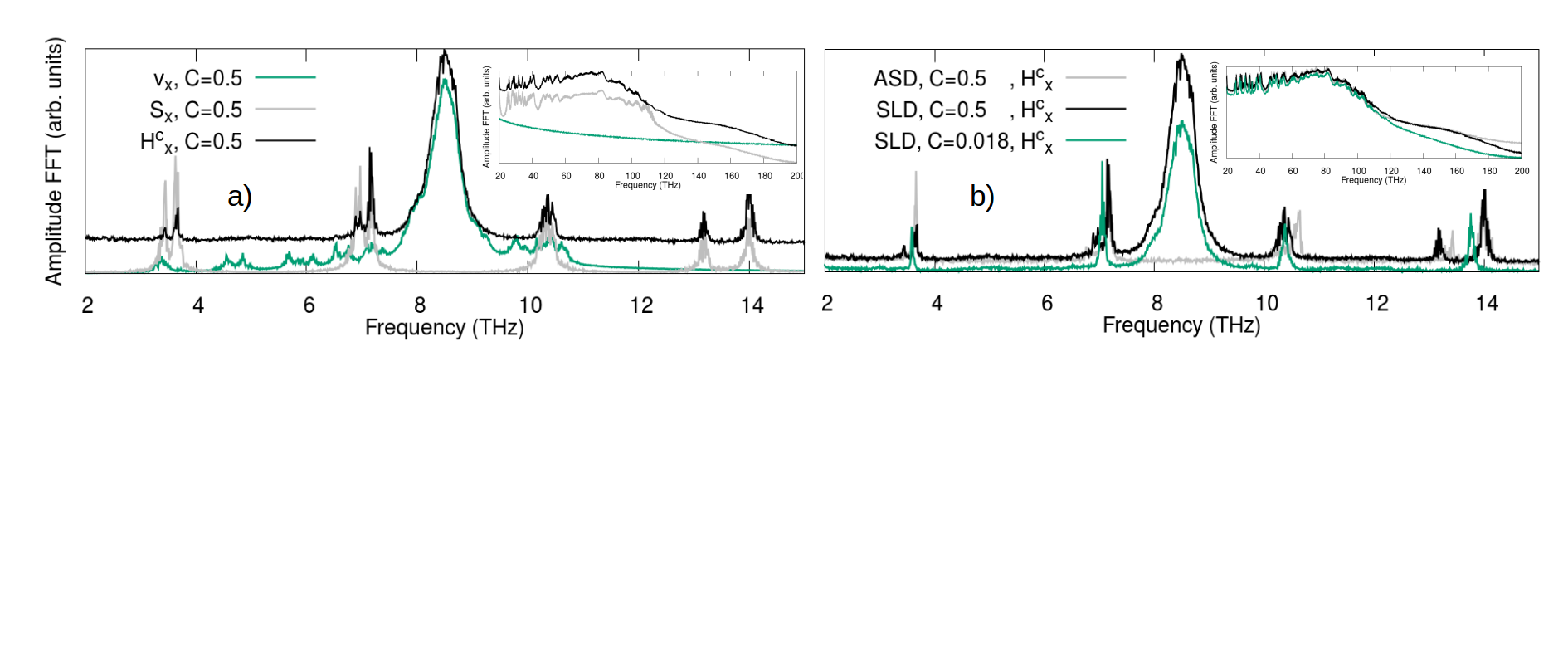}
 \caption{The power spectral density of the auto-correlation function in the frequency domain for magnons, phonons and coupling field for a SLD simulations with a Harmonic lattice, calculated by method $ii)$. Panel a) shows the power density of the auto-correlation function of the x component of the velocity $v_x$, spin $S_x$ and coupling field $H_x^c$. Panel b) presents the power density of the auto-correlation function for the x component of the coupling field for either static (ASD) or dynamic (SLD) lattice. The insets show the high-frequency spectra. For Panel a) the velocity and the coupling field have been multiplied by a factor of 0.12 and 0.05 respectively for easier graphical comparison. } {\label{autocorrelation_combined}}
\end{figure*}

The second method ($ii)$) to investigate the properties of the system involves calculating temporal Fourier transform of individual atoms datasets, and averaging the Fourier response over 1000 atoms of the system. This response represents an integrated response over the k-space. Hence, the projection of intensities on the frequency space presented by method $i)$ has similar features as the spectra presented by method $ii)$. For the results presented in Fig. \ref{autocorrelation_combined}, a system of $10 \times 10 \times 10$ \textsc{bcc} unit cells has been chosen. The system has been equilibrated for a total time of 20 ps with the method presented in $i)$ and the Fast Fourier transform (FFT) is computed for the following 100 ps.

Fig.~\ref{phonons_spectra} shows the phonon spectra for a SLD simulations at $T=300$K, $C=0.5$ for the Morse Potential - Fig.~\ref{phonons_spectra}(a) and the Harmonic Potential - Fig.~\ref{phonons_spectra}(b) calculated for the high symmetry path of a \textsc{bcc} system with respect to both energy and frequency units. The interaction cutoff for both Morse and Harmonic potential is $r_c=\SI{7.8}{\AA}$.
The Morse phonon spectrum agrees well with the spectrum observed experimentally \cite{minkiewicz1967phonon} and  with the results from \cite{ellis2015simulations}. The projection of the spectra onto the frequency domain shows a peak close to 10.5 THz, due to the overlap of multiple phonon branches at that frequency and consequently a large spectral density with many $k$-points excited at this frequency. Moving now to the harmonic potential, parameterised as in Ref. \onlinecite{Assmann2019}, we first note that  
we observe that some of the phonon branches overlap - Fig.\ref{phonons_spectra}b).  Secondly, the projection of intensity onto the frequency domain shows a large peak at 8.6THz, due to a flat region in the phonon spectra producing even larger number of $k$-points in the spectrum which contribute to this frequency.  Finally, the large cutoff makes the Harmonic potential stiffer as all interactions are defined by the same energy, $V_0$, and their equilibrium positions corresponding to a \textsc{bcc} structure. This is not the case for the Morse potential which depends exponentially on the difference between the inter-atomic distance and a constant equilibrium distance, $r_0$.  For a long interaction range, the harmonic approximation will result in a more stiff lattice than the Morse parameterisation.

In principle,  the harmonic potential  with a decreased  interaction cutoff and an increased   strength could better  reproduce the full phonon spectra symmetry for \textsc{bcc} Fe. However, in this work we preferred to use the parameterisation existing in literature~\cite{Assmann2019} and a large interaction cutoff for stability purposes. Although the full symmetry of the \textsc{bcc} Fe phonon spectra is not reproduced by this harmonic potential, the phonon energies/frequencies are comparable to the values obtained with the Morse potential. Nevertheless, we observed the same equilibrium magnetisation and damping (discussed later) for both potentials, hence the simple harmonic potential represents a suitable approximation, that has the advantage of being more computationally efficient.

Fig.~\ref{magnons_morse_spectra} shows the magnon spectrum obtained within the SLD framework using the Morse potential together with its projection onto the frequency domain. The results agree very well with previous calculations of magnon spectra~\cite{perera2017collective,Hellsvik2019}. For the harmonic potential the magnon spectrum is found to be identical to that for the Morse potential with only very small changes regarding the projection of intensity onto the frequency domain. This is in line with our discussion in the previous section where the choice of inter-atomic potential had little effect on the Curie temperature, which is closely linked to the magnonic properties. As the harmonic potential is more computationally efficient than the Morse, we next analyse the properties of the system for a $10 \times 10 \times 10$ unit cells system at $T=300$K with the harmonic potential.

The power spectral density (auto-correlation in Fourier space) of the magnon, phonons and coupling field at \SI{300}{K} is shown in Fig.~\ref{autocorrelation_combined} computed using method $ii$ detailed previously. The amplitude of the FFT spectra of velocities and coupling field has been scaled by 0.12 and 0.05 respectively to allow for an easier comparison between these quantities. As shown in Fig. \ref{autocorrelation_combined}.a) the coupling term presents both magnon and phonon characteristics; demonstrating an efficient coupling of the two sub-systems. The large  peak observed at a frequency of \SI{8.6}{THz} appears as a consequence of the flat phonon spectrum for a Harmonic potential, as observed in the spectrum and its projection onto the frequency domain in Fig. \ref{phonons_spectra}.b). Additionally, Fig. \ref{autocorrelation_combined}.a) can give us an insight into the induced spin noise within the SLD framework. The background of the FFT of the coupling field is flat for the frequencies plotted here, showing that the noise that acts on the spin is uncorrelated. The inset shows a larger  frequency domain where it is clear that  there are no phonon modes for these frequencies, and only thermal noise decaying with frequency is visible. At the same time an excitation of spin modes are visible for frequencies up to ca .100 THz.

 The characteristics of the coupling field with respect to the coupling strength for a dynamic (SLD) and fixed lattice simulations (ASD) are presented in Fig. \ref{autocorrelation_combined}(b).  The only difference between the ASD and SLD simulations is given by the presence of phonons (lattice fluctuations) in the latter.  Since the large peak at \SI{8.6}{THz} is due to the lattice vibrations, it is not present in the ASD simulations. The smaller peaks are present in both models since they are proper magnonic modes. With increasing coupling the width of the peaks increases suggesting that the magnon-phonon damping has increased.  Moving towards the larger frequency regimes, Fig. \ref{autocorrelation_combined}.b) - (inset), we observe that large coupling gives rise to a plateau in the spectra at around \SI{150}{THz}, which is present as well for the fixed-lattice simulations (ASD). The plateau arises from a weak antiferromagnetic exchange that appears at large distances due to the competition between the ferromagnetic exchange and the antiferromagnetic exchange-like term in the pseudo-dipolar coupling. 

  \begin{figure*}[t]\centering     
 \includegraphics[width=2\columnwidth, trim={2cm 10cm 2cm 0}]{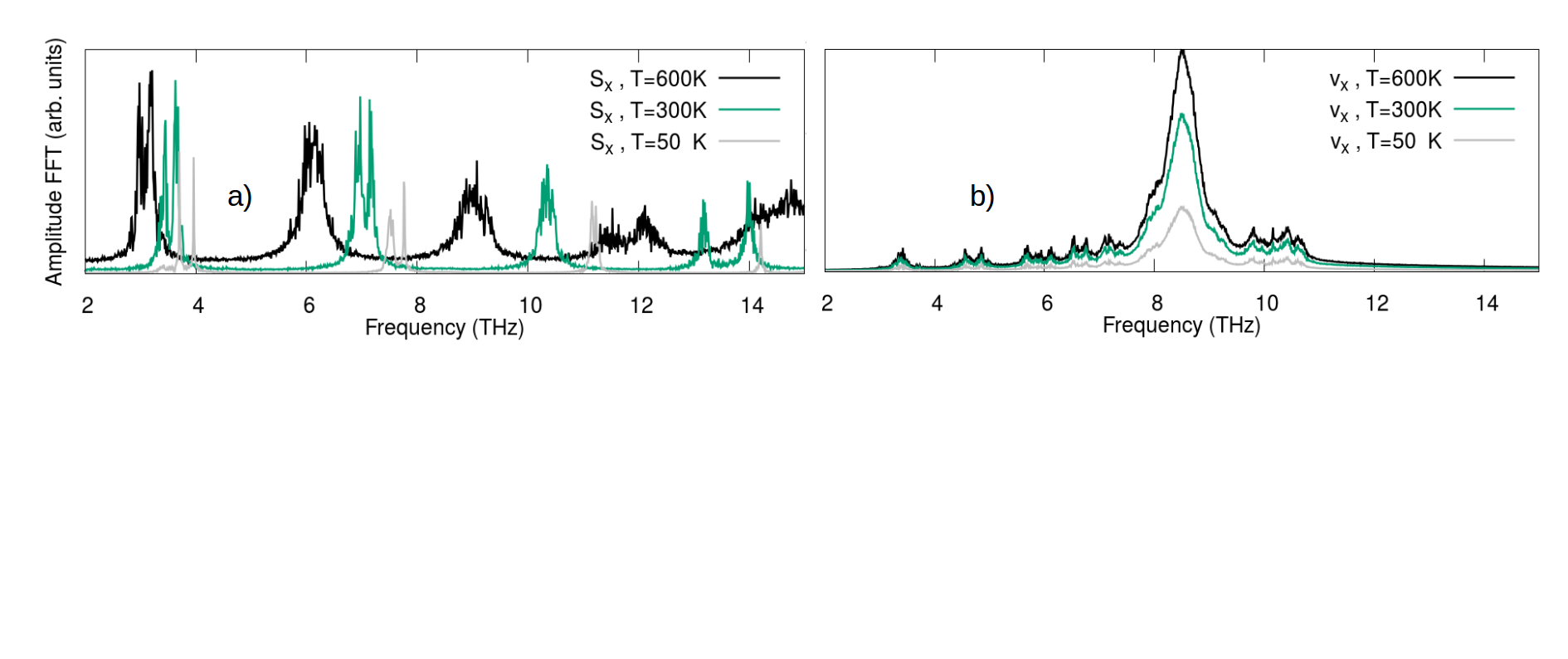}
 \caption{The power spectral density of the auto-correlation function in the frequency domain for magnons - Panel a) and phonons - Panel b) for a SLD simulations with a Harmonic lattice, calculated by method $ii)$, for three distinct temperatures and a coupling constant of  $C=0.5 $.  } {\label{autocorrelation_temp}}
\end{figure*}

 We have also analysed the characteristics of the magnon and phonon spectra for different temperatures- Fig. \ref{autocorrelation_temp}.
  With increasing temperature, the peaks corresponding to magnons shift to smaller frequencies. This is a typical situation known as a softening of low-frequency magnon modes due to the influence of thermal population, see e.g.\cite{Atxitia2010}  
   - Panel a). The same effect can be observed by calculating the magnon spectra via method \textit{i} for various temperatures. In Panel b), the peak corresponding to phonons remains almost at the same frequency of about \SI{8.6}{THz}, as the phonon spectra is not largely affected by temperatures up to $T=600$K. The increase of the effective damping (larger broadening) of each magnon mode with temperature is clearly observed.

\section{Macrosopic magnetisation damping}
\label{sec::magnon_phonon_damping}

\begin{figure*}[!htb]\centering   
 \includegraphics[width=1.5\columnwidth, trim={0cm 1cm 0 0cm}]{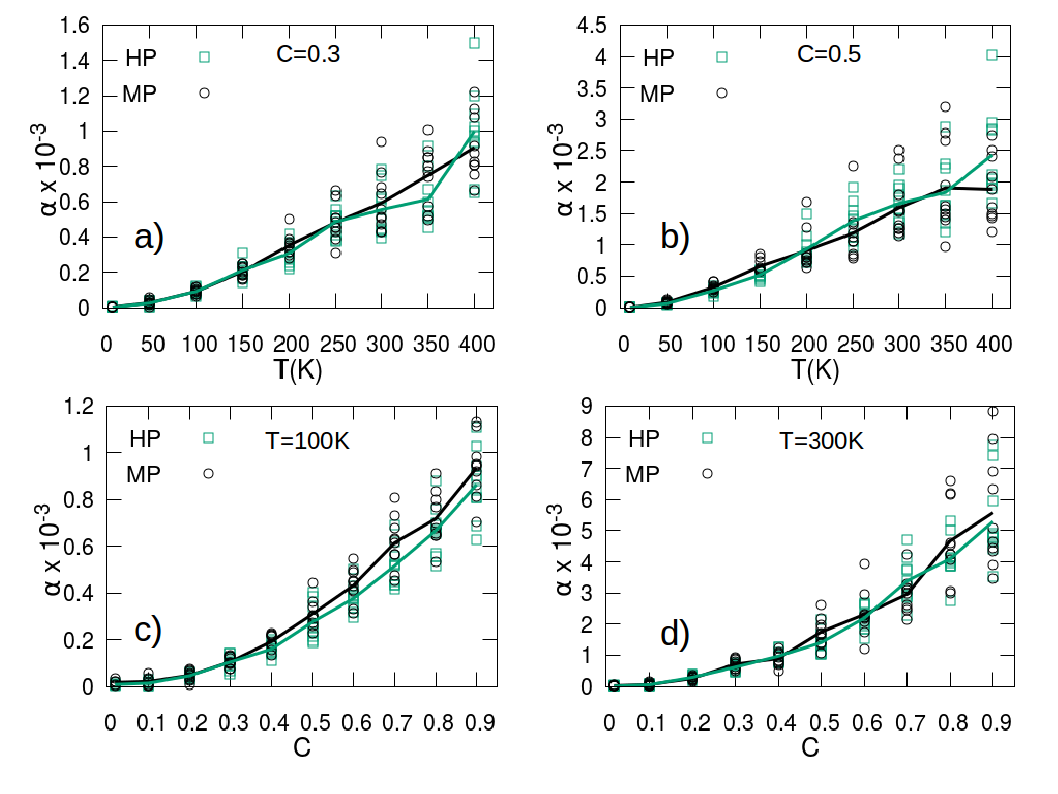}
 \caption{ Damping parameter extracted from fitting the z component of the magnetisation for two different choices of potential: HP- Harmonic Potential (green open squares) and MP-Morse Potential (black open circles) as function of temperature Fig. a), b) and as function of the coupling strength Fig. c), d); Fig a) and b) are calculated for a constant coupling strength of $C=0.3$, $C=0.5$ respectively. Fig c) and d) are calculated for  temperatures of $T=100K$, $T=300K$ respectively. The black and green lines represents the average damping parameter obtained from the simulations using the Morse  and the Harmonic Potentials, respectively. }{\label{damping_all}}
\end{figure*}

In this section we evaluate the macroscopic damping parameter experienced by magnetisation due to the magnon-phonon excitations for a periodic \textsc{bcc} system using our SLD model. This method for calculating the damping has been presented in \cite{chubykalo2006, ellis2012classical, strungaru2020}. The system is first thermalised at a non-zero temperature in an external field of $B_{ext} = 50$T applied in the $z$ direction, then the magnetisation is rotated coherently through an angle of $30^ \circ$. The system then relaxes back to equilibrium  allowing the relaxation time to be extracted. The averaged $z$ component of magnetisation is then fitted to the function $m_z(t)=\tanh(\alpha \gamma B_{ext} (t + t_0)/(1+\alpha^2))$ where $\alpha$ represents the macrosopic (LLG-like) damping, $\gamma$ the gyromagnetic ratio and $t_0$ a constant related to the initial conditions. The model system consists of $10 \times 10 \times 10 $ unit cells and the damping value obtained from fitting of $m_z(t)$  is averaged over 10 different simulations.

Fig.~\ref{damping_all} shows the dependence of the average damping parameter together with the values obtained from individual simulations for different temperatures and coupling strengths for two choices of mechanical potential. In our model, the spin system is thermalised by the phonon thermostat, hence no electronic damping is present.  With increasing coupling, the energy and angular momentum transfer is more efficient, hence the damping is enhanced.  Since the observed value of induced damping is small, calculating the damping at higher temperature is challenging due to the strong thermal fluctuations that  affect the accuracy of the results. Despite the low temperatures simulated here, the obtained damping values  (at $T=50$K, $\alpha=4.9\times10^{-5}$) are of the same order as reported  for magnetic insulators such as YIG ($1\times10^{-4}$ to $1\times10^{-6}$ \cite{jermain2017increased, Maier-Flaig2017} ) as well as in different SLD simulations ($3\times10^{-5}$, \cite{Assmann2019}). Generally, the induced  damping value depends on the phonon characteristics and the  coupling term, that allows transfer of both energy and angular momentum between the two subsystems.

\begin{figure}[tb]\centering   
 \includegraphics[width=0.9\columnwidth, trim=0.5in 0.25in 0.25in 0in]{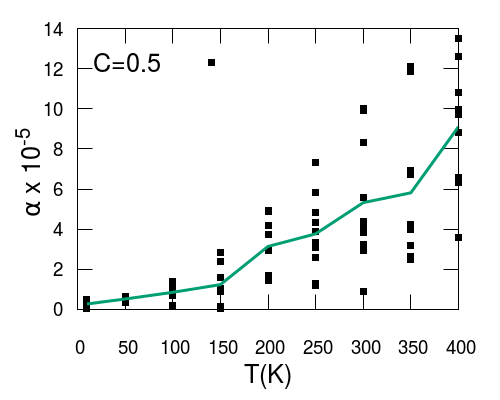}
 \caption{ Temperature variation of the damping parameter for N\'eel-like on-site coupling, $H_c= -\sum_{i,j} f({r}_{ij}) ((\textbf{S}_i \cdot \hat {\textbf{r}}_{ij})^2 - \frac {1}{3} \textbf{S}_i^2)$. The values are extracted from $m_z(t)$ fittings for 10 realisations; }{\label{damping_neel}}
\end{figure}

 Fig.~\ref{damping_all}(a) and (b) compare the calculated damping for the Morse and Harmonic potential for two values of the coupling strength. We observe that the values are not greatly affected by the choice of potential. This arises due to the fact that only the spin modes around $\Gamma$ point are excited and for this low k-vectors modes the inter-atomic distances between neighbouring atoms do not vary significantly. The extracted  damping parameter as a function of coupling strength for \SI{100}{K} and \SI{300}{K } is presented in Fig. \ref{damping_all}(c) and (d) respectively. The functional form of the variation is quadratic, in accordance with the form of the coupling term. Measurements of damping in magnetic insulators, such as YIG, show a linear increase in the damping with temperature, \cite{Maier-Flaig2017} which agrees with the relaxation rates calculated by Kasuya and LeCraw \cite{kasuya1961relaxation} and the relaxation rates calculated in the NVE SLD simulations in Ref.~\onlinecite{Assmann2019}.
 However, Kasuya and LeCraw suggest that the relaxation rate can vary as $T^n$, where $n=1-2$ with $n=2$ corresponding to larger temperature regimes. Nevertheless, the difference between the quadratic temperature variation of the damping observed in our simulations and the linear one observed in experiments for YIG can be attributed to the difference in complexity between the \textsc{bcc} Fe model and YIG.
 The difference between the trends may as well suggest that the spin-orbit coupling in YIG could be described better by a linear phenomenological coupling term, such as the one used in Refs.~\onlinecite{Perera2016} and \onlinecite{Karakurt2007}, but we note that such forms can lead to a uniform force in the direction of the magnetisation and so might need further adaptation before being suitable. To test an alternate form of the coupling we have changed the pseudo-dipolar coupling to an on-site form, specifically $\mathcal{H}_c= -\sum_{i,j} f({r}_{ij}) ((\textbf{S}_i \cdot \hat {\textbf{r}}_{ij})^2 - \frac {1}{3} \textbf{S}_i^2)$ i.e a N\'eel-like anisotropy term. This leads to much smaller damping as shown in Fig. \ref{damping_neel} ($T=300$~K, $\alpha=3.3\times10^{-5}$, averaged over 5 realisations) making it difficult to accurately calculate the temperature dependence of the damping, especially for large temperatures. The magnon-phonon damping can clearly have complex behavior depending on the properties of the system, especially the coupling term, hence no universal behaviour of damping as function of temperature can be deduced for spin-lattice models. 

Neglecting the lattice contribution, the temperature dependence of the macrosopic damping can be mapped onto the Landau-Lifshitz-Bloch formalism (LLB)\cite{chubykalo2006} and theory \cite{Nieves2014} and ASD simulations \cite{Ellis2019} have shown it to vary inversely with the equilibrium magnetisation. The LLB theory shows that the macrosopic damping is enhanced for large temperatures due to thermal spin fluctuations. Using the equilibrium magnetisation it is possible to approximate the variation of damping  with temperature produced  due to thermal fluctuations within the LLB model. From 100K to 400K the damping calculated via the  LLB model increases within the order of $10^{-5}$, which is considerably smaller than the results obtained via the SLD model. This shows that within the SLD model the temperature increase of the damping parameter is predominantly due to magnon-phonon interaction, and not due to thermal magnon scattering, as this process is predominant at larger temperatures.

\section{Conclusions and outlook}
To summarise, we have developed a SLD model that is able to  transfer energy and angular momentum efficiently from the spin to lattice sub-systems and vice-versa via a pseudo-dipolar coupling term. Our approached takes the best features from several previously suggested models and generalize them which allows modelling in both canonical and microcanonical ensembles.  With only the lattice coupled to a thermal reservoir and not the spin system, we  reproduce the temperature dependence of the equilibrium magnetisation, which agrees with the fact that the spin-lattice model obeys the fluctuation-dissipation theorem. We are able to study the dynamic properties such as phonon and spin spectrum  and macrosopic damping, showing that the magnetic damping is not greatly influenced by the choice of potential, however it is influenced by the form of the coupling term. This enables the possibility of tailoring the form of the coupling term so it can reproduce experimental dependencies of damping for different materials. In future, the addition of quantum statistics for Spin Lattice Dynamics models\cite{EvansPRB2015,BarkerPRB2019}  may also yield better agreement with experimental data.

The SLD model developed here opens the possibility of the investigation of ultrafast dynamics experiments and theoretically studies of the mechanism through which angular momentum can be transferred from spin to the lattice at ultrafast timescales. As we have demonstrated that the model works well in the absence of an phenomenological Gilbert damping, which consists mainly of electronic contributions, the SLD model can be employed to study magnetic insulators, such as YIG, where the principal contribution to damping is via magnon-phonon interactions.  Future application of this model includes controlling the magnetisation via THz phonons ~\cite{Melnikov2003} which can lead to non-dissipative switching of the magnetisation ~\cite{Vlasov2020,Kovalenko2013}. With the increased volume of data stored, field-free, heat-free switching of magnetic bits could represent the future of energy efficient recording media applications. Another possible application is more advanced modelling of the ultrafast Einstein-de-Haas effect ~\cite{Dornes2019} or phonon-spin transport ~\cite{Ruckriegel2020}.

\section{Acknowledgements}
We are grateful to Dr. Pui-Wai Ma and Prof. Matt Probert for helpful discussions. Financial support of the Advanced Storage Research Consortium is gratefully acknowledged. MOAE gratefully acknowledges support in part from EPSRC through grant EP/S009647/1.
The spin-lattice simulations were undertaken on the VIKING cluster, which is a high performance compute facility provided by the University of York. The authors acknowledge the networking opportunities provided by the European COST Action CA17123 "Magnetofon" and  the short-time scientific mission awarded to  M.S.

\bibliography{references}

\end{document}